\newcommand{\Mpc}{h$^{-1}$Mpc}
\newcommand{\kms}{km s$^{-1}$}

\documentstyle[12pt,aaspp4]{article}

\begin{document}

\title{Superclustering at Redshift ${\mathbf z=0.54}$}

\author{A. J. Connolly\altaffilmark{1} and A. S. Szalay\altaffilmark{1,2}}
\affil{Department of Physics and Astronomy, The Johns Hopkins
University, Baltimore, MD 21218}

\author{D. Koo\altaffilmark{1}}
\affil{University of California Observatories, Lick Observatory, 
Board of Astronomy and Astrophysics, University of California Santa
Cruz, CA 95064}

\author{A. K. Romer}
\affil{Dept. of Physics and Astronomy, Northwestern University, 2145 Sheridan 
Road, Evanston, Il-60208, USA.} 

\author{B. Holden and R.C. Nichol}
\affil{Department of Astronomy and Astrophysics, University of Chicago, 
5460 S. Ellis Ave, Chicago, IL 60637}

\author{T. Miyaji}
\affil{ Max-Planck-Institut f\"ur Extraterrestrische Physik
                     Postf. 1603, D-85740, Garching, Germany
}
\altaffiltext{1} {Visiting Astronomer, Kitt Peak National Observatory,
National Optical Astronomy Observatories, operated by the Association
of Universities for Research in Astronomy, Inc., under contract with
the National Science Foundation.}

\altaffiltext{2}{Department of Physics, E\"{o}tv\"{o}s University, 
Budapest, Hungary, H-1088}

\clearpage

\begin{abstract} 

We present strong evidence for the existence of a supercluster at a
redshift of $z=0.54$ in the direction of Selected Area 68. From
the distribution of galaxies with spectroscopic redshifts we find that
there is a large over-density of galaxies (a factor of four over the
number expected in an unclustered universe) within the redshift range
$0.530 < z < 0.555 $. By considering the spatial distribution of
galaxies within this redshift range (using spectroscopic and
photometric redshifts) we show that the galaxies in SA68 form a linear
structure passing from the South-West of the survey field through to
the North-East (with a position angle of approximately 35$^\circ$ East
of North). This position angle is coincident with the positions of the
X-ray clusters CL0016+16, RX J0018.3+1618 and a new X-ray cluster, RX
J0018.8+1602, centered near the radio source 54W084.  All three of
these sources are at a redshift of $z \sim 0.54$ and have position
angles, derived from their X-ray photon distributions, consistent with
that measured for the supercluster.  Assuming a redshift of 0.54 for
the distribution of galaxies and a FWHM dispersion in redshift of
0.020 this represents a coherent structure with a radial extent of 31
h$^{-1}$Mpc, transverse dimension of 12 \Mpc, and a thickness of
$\sim$ 4 \Mpc. The detection of this possible supercluster
demonstrates the power of using X-ray observations, combined with
multicolor observations, to map the large scale distribution of
galaxies at intermediate redshifts.

\end{abstract}


\section{Introduction}

By mapping the spatial distribution of galaxies, we can determine the
intrinsic scales on which galaxies cluster, from poor groups through
rich clusters to superclusters. Quantifying the abundance of these
clusterings and their evolution with time should provide important
constraints on the multifarious cosmological models (Liddle et al.\
1996). In the local universe, extensive redshift surveys have been
undertaken to map the distribution of galaxies to $B \sim 15.5$ that
reach to a redshift of $z \sim 0.05$. Such surveys have uncovered
coherent structures that extend over many tens of megaparsecs, such as
the Perseus-Pisces filament (Giovanelli and Haynes 1993) or the
sheet-like ``Great Wall'' (Geller and Huchra 1989). These structures
appear to be common in other and even deeper ($B < 20$) surveys that
extend beyond redshifts $z \sim 0.1$ (Landy et al.\ 1996, Willmer et
al.\ 1996).

At even fainter magnitudes, redshift surveys have been undertaken to a
limit of $B\simeq 24$ (Cowie et al. 1996, Glazebrook et al.\ 1995,
Lilly et al.\ 1995). Because of the long exposures required to reach
such faint limits, these surveys have been restricted to a few
pointings that cover tiny regions of sky, typically $\le$ 10 arcmin in
diameter. Consequently, while large over-densities of galaxies are
visible in the deep redshift surveys (Broadhurst et al.\ 1990, Le Fevre
et al.\ 1994, Cohen et al.\ 1996), the angular distribution of these
structures and, therefore, their transverse spatial extent, have yet
to be well studied.

In this letter we consider the distribution of galaxies in the
direction of Selected Area 68 (Kron 1980). We identify this region as
a potential site of an intermediate redshift supercluster because of
the coherent distribution of very red galaxies, estimated to be at $z
\sim 0.5$, found by Koo (1985) and the presence of two X-ray clusters
at the same redshift (Hughes et al.\ 1995).  The over-density of
galaxies is quantified using a small sample of spectroscopic redshifts
in $\S2$ and then improved using a larger sample of photometric
redshifts in $\S3$. The coincidence of this structure with the
position and orientation of three X-ray clusters is discussed in
$\S4$. Finally we show that the galaxies and clusters are consistent
with a ``Great Wall''-like supercluster, at $z=0.54$, that is almost
edge-on with respect to the line of sight.  For this letter we assume
an H$_o = 100$ \kms\ Mpc$^{-1}$ and $\Omega=1$ cosmology; at redshift
$z \sim 0.54$, the lookback time is nearly half the age of the
universe and one degree spans $13.5$ \Mpc.

\section{The Distribution of Spectroscopic Redshifts in the 
Selected Area SA68}

To determine if there exist intermediate redshift counterparts to the
locally detected superclusters, we consider the spectroscopic and
photometric galaxy catalog of Koo and Kron (Kron 1980, Koo 1986, Munn
et al.\ 1996) in the direction of Selected Area 68 (00$^{\rm
hr}$14$^{\rm m}$ +15$^\circ$30$^{\rm m}$). These data comprise a total
of 1750 galaxies, $B_J <23.0$, with multicolor photometry in the $U,
B_J, R_F$ and $I_N$ passbands. Of these galaxies 286 have
spectroscopically measured redshifts.

In Fig 1 we show the redshift distribution of the spectroscopic sample
of galaxies. The most striking features of this redshift distribution
are the regular set of peaks. Such over-densities have been interpreted
as the intersection of the narrow pencil beam surveys with large-scale
coherent structures in the galaxy distribution (Broadhurst et al.\
1990, Szalay et al.\ 1991, Le Fevre et al.\ 1994, Cohen et al.\
1996). For later discussions, we consider only the spatial and
redshift distribution of those galaxies lying between $0.530 < z <
0.555$.

To test whether this feature, at $z=0.55$, represents a significant
perturbation from number of galaxies expected from an unclustered
distribution of galaxies, we calculate the selection function for the
spectroscopic sample of galaxies. As the spectroscopic sample is not
formally magnitude limited we must calculate the expected redshift
distribution using the models of Gronwall and Koo (1995). These models
were derived to match the observed redshift distribution of SA68 to a
limit of B$<24$. Taking the observed $B_J$ magnitude distribution of
the spectroscopic galaxy sample we construct the expected {\it dn/dz}
(the solid line in Fig 1). The expectation value for the number of
galaxies with $0.530 < z < 0.555$ is 3.54.

To account for a clustered distribution we calculate the variance of
the expectation value by integrating the local spatial correlation
function, $\xi (r)$, over this redshift volume. Assuming a projected
area of 706 arcmin$^2$ for SA68 we estimate the variance in the
expectation value to be 3.7. It should be noted that this is a
conservative value (an upper limit) as we assume no evolution of the
spatial correlation function, with redshift.  We would, therefore,
expect to detect
\begin{equation}
<N> = 3.54 \pm 3.7
\end{equation}
galaxies with spectroscopic redshifts within the redshift range $0.530
< z < 0.555$.  The observed number of galaxies in this redshift peak
is 14, representing a factor of 4 over the unclustered expectation
value. Fitting a Gaussian to the spectroscopic redshifts we determine
the dispersion of the redshift distribution to be 0.0083. If the
dispersion in redshift were due to the internal velocity of a cluster
this would equate to a velocity dispersion of $\sim 2500$ \kms\ far in
excess of local observations (Zabludoff et al.\ 1993, Collins et al.\
1995). The radial distance corresponding to the FWHM of the redshift
distribution is then 31 \Mpc.

The projected angular distribution of the spectroscopic sample of
galaxies within this redshift range is shown in Fig 2a.  They form a
linear structure passing from the South-West of the SA68 field through
to the North-East. The angular distribution extends to the limits of
the survey field. From the second moments of the galaxy distribution
we find that this structure can be represented by an ellipse with a
position angle oriented 29$^\circ$ East of North and centered on
00$^{\rm hr}$14$^{\rm m}$40+15$^{\circ}$33$^{\rm m}$58 (B1950).

While the number of galaxies with observed spectroscopic redshifts is
small we can estimate whether they are consistent with a uniform
distribution of galaxies across the SA68 survey field. We do this by
applying a 2 dimensional Kolmogorov-Smirnov test (Peacock 1983, Fasano
and Franceschini 1987). For a sample size of 14, the probability that
the galaxies with spectroscopic redshifts are consistent with a
uniform distribution is only 15\%.

\section{The Distribution of Galaxies across Selected Area SA68}

Since the spectroscopic sample is so small, we adopt another technique
to increase the redshift sample, namely we estimate the redshift of a
galaxy from its broadband magnitudes. The effectiveness of this
technique for deriving galaxy redshifts to the limit of our
photometric data has been demonstrated by Connolly et al.\ (1995). By
fitting the $U, B_J, R_F, I_N$ magnitudes with a second order relation
to the spectroscopic redshifts, redshifts can be estimated to an
accuracy of $\sigma_z \le 0.05$.

We calculate the photometric redshift relation for the galaxies in the
photometric sample using the prescription given by Connolly et al.\
(1995). Comparing those galaxies with spectroscopic redshifts with
their estimated photometric redshifts we derive a dispersion in the
relation of $\sigma_z = 0.049$. We select a 1 $\sigma_z$ range around
the peak in the spectroscopic redshift distribution. In Fig 2b we show
the distribution of the 146 galaxies within the redshift range $0.49 <
z < 0.59$. These data have been smoothed with a kernel of diameter 2.5
arcmin (0.55 h$^{-1}$Mpc at $z=0.54$; equivalent to the core diameter
of a King cluster profile).

The galaxy distribution is again seen to form a coherent linear
structure passing from the South-West to the North-East of SA68. The
second moments of the distribution of galaxies gives a position angle
of 40.3$^\circ$ centered on 00$^{\rm hr}$14$^{\rm
m}$41+15$^{\circ}$37$^{\rm m}$16. Clearly this represents an
underestimate of the over-density of galaxies within the supercluster
(due to the contamination from foreground and background galaxies).
The two-dimensional Kolomogorov-Smirnov test yields only a 2\%
probability that the photometric-redshift sample is drawn from a
uniform distribution. A comparison of the angular distributions of
those galaxies with spectroscopic redshifts with those from the
photometric redshift sample shows that the probability that they are
drawn from {\it different} intrinsic populations is only 45\% (i.e.\
less than a one sigma deviation). The angular distribution of the
spectroscopic and photometric redshift samples, therefore, display the
same large-scale clustering properties. They show a linear structure
of at least 0.5 degrees in extent (equivalent to 6.5 \Mpc\ at
$z=0.54$).

\section{The distribution of X-ray clusters around Selected Area SA68}

Just beyond half a degree from the center of SA68 lies the X-ray
luminous cluster CL0016+16 at a redshift of $z=0.5455$ (Koo 1981, see
Fig 2), which was the target of one of the deepest ROSAT PSPC pointed
observations (Hughes et al.\ 1995). The efficacy of using deep ROSAT
pointings to serendipitously identify X-ray clusters has been
demonstrated by Hughes et al.\ (1995). As part of the Serendipitous
High--redshift Archival ROSAT Cluster (SHARC) survey (Nichol et al.\
1996, Burke et al.\ 1996) we have, therefore, reanalyzed this deep
pointing using a detection algorithm based on the wavelet transform.

In the SHARC survey, an X-ray source is flagged as a candidate distant
cluster if the observed X-ray emission is significantly extended ($> 3
\sigma$) compared to the radial--dependent PSPC point-spread
function. Furthermore, the source is required to have no optical
counterpart on the Palomar Digital Sky Survey Plates. In addition to
CL0016+16 (00$^{\rm hr}$18$^{\rm m }$33.2+16$^{\circ}$26$^{\rm m}$18)
and RX J0018.3+1618 (00$^{\rm hr}$18$^{\rm m
}$16.8+16$^{\circ}$17$^{\rm m}$45, Hughes et al.\ 1995) - which both
satisfy these criteria - we have discovered a further such X-ray
source; RX J0018.8+1602 (00$^{\rm hr}$18$^{\rm m
}$45.5+16$^{\circ}$01$^{\rm m}$41). This source lies 25 arcmins from
the center of the PSPC pointing and is within 1 arcmin of the radio
galaxy 54W084 (Neff et al.\ 1995).

The PSPC X-ray contours of RX J0018.8+1602 are overlaid on a B$_J$
photographic plate (\# 1286) observed by R. Kron with the KPNO 4m
telescope. The peak in the X-ray photon distribution is coincident
with an over-density of faint, B$_J <23$, galaxies in the optical
data. The color of the central optical galaxy is consistent with that
of an elliptical galaxy at a redshift of 0.5. Within the X-ray
contours lies the radio galaxy 54W084 at a redshift of $z=0.544$
(R. Windhorst, private communication); indicated by an arrow in Fig
3. If we assume a redshift of 0.544, RX J0018.8+1602 has a luminosity
in the 0.500--2.0 keV energy range of 4.25$\times 10^{43}$ ergs
s$^{-1}$. This compares with an X-ray luminosity of 2.5$\times
10^{43}$ ergs s$^{-1}$ measured by Hughes et al.\ (1995) for the
cluster RX J0018.3+1618. The ROSAT PSPC spectrum of RX J0018.3+1618 is
consistent with a thermal plasma. Assuming a Raymond-Smith model with
heavy metal abundance of 0.3 solar and the Galactic N$_H$ value at the
cluster position, the temperature of the plasma is
$kT=1.6^{+0.7}_{-0.4}{\rm keV}$. This source has also been detected by
ASCA and a preliminary analysis shows a spectrum consistent that
derived from the ROSAT data. This is indicative of an intermediate
redshift X-ray cool cluster of galaxies.

The redshifts of CL0016+16 and RX J0018.3+1618 are 0.5455 and 0.5506
respectively. All three X-ray clusters are, therefore, consistent with
the redshift distribution observed in SA68.  CL0016+16 and RX
J0018.3+1618 have been suggested to be a bound system and possibly
linked with the over-density of red galaxies in SA68 found by Koo
(Hughes et al.\ 1995). Below we show in fact that the space
distribution of the three clusters and those galaxies in SA68 with
spectroscopic redshifts are consistent with a supercluster viewed edge
on with respect to the line of sight.

The orientation of clusters of galaxies has been suggested as a means
of identifying coherent large scale structures (Bingelli 1982).
Results derived from the optical distribution of galaxies remain
inconclusive with West (1989) finding evidence for a correlation
between cluster position angles on scales of up to 45 \Mpc\ while
Struble and Peebles (1985), using similar data, find no significant
signal. Much of these uncertainties arise from the difficulty in
separating cluster galaxies from contamination due to background
sources (especially in the outer regions of a cluster). Determining
the orientation of a cluster from the distribution of its X-ray
emitting gas provides a more objective measure (Ulmer et al.\ 1989)
since the gas better traces the cluster potential.

The position angles of the three X-ray sources were determined from
the second moments of the outer isophotes of the X-ray photon
distribution. The derived values for CL0016+16, RX J0018.3+1618 and RX
J0018.8+1602 were 39$^\circ$, 29$^\circ$ and 30$^\circ$ East of North
respectively. All three sources are, therefore, aligned with the
galaxy distribution of SA68. As has been noted by Bond et al.\ (1996)
the signature of superclustering will be strongest when the individual
clusters are aligned.

\section{Discussion: A ``Great Wall'' at ${\mathbf z=0.54}$}

Combining the optical and X-ray data we have strong evidence for a
coherent structure, at a redshift of $z=0.54$, extending about one
degree across the sky from the survey field SA68 through the cluster
CL0016+16. At this redshift, this translates to a tangential size of
12 \Mpc, a radial depth of 31 \Mpc and a ``thickness'' of 4 \Mpc. The
positions of the galaxies and clusters within this volume are not
randomly distributed but appear to lie in a planar distribution (i.e.\
their redshifts and angular distribution are strongly correlated).

To determine the true geometry of the galaxy distribution, i.e.\
whether it is better represented by an extended filament or a sheet of
galaxies, we fit a two dimensional surface to the spectroscopic
redshifts. The redshifts of the galaxies in SA68 and the three
clusters are transformed to comoving distance and treated as
independent points, we do not weight the cluster redshifts by the
number of galaxies that have spectroscopic observations. The best fit
to these data is a plane with an orientation 40$^\circ \pm 10^\circ$
East of North and an angle 12$^\circ \pm 2^\circ$ from the line of
sight. Given that the redshift dispersion of the galaxies exceeds that
expected for a cluster, we suggest that the structure we are observing
is a sheet of galaxies oriented almost orthogonally to our line of
sight.

As noted in 3.\ the dispersion in the photometric-redshift relation
results in a dilution of the supercluster due to contamination by
foreground and background galaxies. We cannot, therefore, map the full
three-dimensional distribution of the supercluster. We can, however,
determine whether the galaxy distribution is comparable to that
observed in the local universe (i.e.\ if a structure similar to the
``Great Wall'' were to exist at $z=0.54$ what would its signature
be). If we correct for the orientation of the galaxy distribution and
determine the width orthogonal to the plane of the supercluster we
derive a dispersion of 433$^{+85}_{-61}$ \kms.  This is consistent
with the mean value of 300 \kms\ determined by Ramella et al.\ (1992)
for the ``Great Wall''. Furthermore, if we assume that 30\% of the
galaxies within the redshift range $0.5 < z <0.6$ are indeed members
of the supercluster, as suggested by the results of de Lapparent et
al.\ (1991) from their analysis of the CfA redshift survey, we can
estimate the surface density of the galaxies in the
supercluster. Allowing for an inclination angle of 10 degrees to the
line of sight and a luminosity distance of 1837 \Mpc\, we estimate the
surface density of galaxies to be approximately 0.6 h$^2$
Mpc$^{-2}$. This is only slightly higher than the values of 0.25--0.4
h$^2$ Mpc$^{-2}$ determined for the individual slices in the CfA
redshift survey (de Lapparent et al.\ 1991). It would appear,
therefore, that we have identified an intermediate redshift
counterpart to the sheet-like supercluster structures observed in
redshift surveys of the local universe.

Clearly the available data, while providing tantalizing evidence for
large scale clustering at intermediate redshifts, do not map the full
extent of this supercluster. The over-density in the spectroscopic
redshifts encompasses the full extent of the survey field SA68. To
determine the tangential distribution of galaxies at a redshift of
0.54 and, therefore, the true size of this supercluster we are engaged
in followup photometric and spectroscopic observations of this region.

\acknowledgments 

We would like to thank the referee for helpful comments that improved
the clarity of this paper.  We thank Richard Kron for providing the
photographic plate for SA68 and help in the identification of the new
X-ray cluster. We are grateful to Kron, Jeffrey Munn, Steven Majewski,
Matthew Bershady, and John Smetanka for pre-publication access to the
KPGRS catalogs and to Rogier Windhorst for the redshift of 54W084.  We
acknowledge Piero Rosati and Jack Hughes for useful discussions on the
X-ray distributions in CL0016+16 and Gerry Luppino, Gordon Richards,
Dan Vanden Berk and Michael Strauss for attempting optical followups
of the SA68 region.  AJC and AS acknowledge partial support from NSF
grant AST-9020380, an NSF-Hungary Exchange Grant, the US-Hungarian
Fund and the Seaver Foundation. DCK acknowledges support from the NSF
grant AST 88-58203.  AKR acknowledges support from NASA grant
NAGS-2432. BH was supported in part by the National Science Foundation
under a cooperative agreement with the Center for Astrophysical
Research in Antarctica (CARA), grant number NSF OPP 89-20223. CARA is
a National Science Foundation Science and Technology Center. TM
appreciates the hospitality of NASA Goddard Space Flight Center and
the Johns Hopkins University during his visit. His visit was partially
supported by USRA.

\clearpage

\begin{figure}
\caption{The redshift distribution of galaxies in the Selected Area
SA68 for those galaxies in the Koo and Kron spectroscopic sample with
$B_J<23$.  The solid line represents the expected distribution of
redshifts for this sample assuming no clustering. The expectation
value for the number of galaxies between $0.530 < z < 0.555$ is $3.54
\pm 3.7$. The observed number of galaxies within this redshift range is
14, a factor of four larger.}
\end{figure}

\begin{figure}

\caption{The spatial distribution of those galaxies with spectroscopic
redshifts between $0.530 < z < 0.555$, in SA68, is shown in Fig
2a. The solid line shows the extent of the SA68 survey field. The
distribution of galaxies with photometric redshifts in the range $0.49
< z < 0.59$ is given in Fig 2b.  The spectroscopic and photometric
samples of galaxies form a linear distribution passing from the
South-West of SA68 through to the North-East with position angles of
27.2$^\circ$ and 40.3$^\circ$ (East of North) respectively. The
positions of the previously identified X-ray clusters CL0016+16 and RX
J0018.3+1618 are shown in both figures together with the new cluster
RX J0018.8+1602.  The position angle of the photon distribution for
these X-ray clusters are 39$^\circ$, 29$^\circ$ and 30$^\circ$
respectively. }
\end{figure}

\begin{figure}

\noindent {\bf Figure 3 can be found as a compressed postscript file (4MB) 
at http://tarkus.pha.jhu.edu/$\sim$ajc/papers/supercluster/Figure3.ps.Z}\\

\caption{The ROSAT PSPC X-ray contours for RX J0018.8+1602 are
overlaid on a 4m B$_J$ photographic plate of SA68. A local bore sight
correction has been applied to the X-ray data based on the position of
three stars. The wavelet analysis of the field shows the X-ray
distribution to be more extended than the radially dependent PSF at
greater than the three sigma level. The peak in the X-ray photon
distribution is coincident with an over-density of faint galaxies
(B$_J$ $\sim$ 23) with the central galaxy appearing elliptical.  The
arrow indicates the position of the radio galaxy 54W084 at a redshift
of $z=0.544$. The position of this radio source is within one arcmin
of the center of the X-ray cluster and is contained within the outer
isophotes. If we assume the radio galaxy is a member of the cluster
the X-ray luminosity corresponds to 4.25$\times 10^{43}$ ergs
s$^{-1}$. This is approximately 13\% of the luminosity of CL0016+16.}
\end{figure}


\newpage

\begin{references}

\reference {} Bingelli, B., 1982, A\&A, 107, 338

\reference {} Bond, J.R., Kofman, L. \& Pogosyan, D., 1996, Nature, 380, 603

\reference {} Broadhurst, T.J., Ellis, R.S., Koo, D.C. \& Szalay, A.S.,
1990, Nature, 343, 726.

\reference {} Burke, D.J., Collins, C.A., Nichol, R.C., Romer, A.K.,
Holden, B. P., Sharples, R. M., Ulmer, M. P.,
Proc. ``R\"{o}ntgenstrahlung from the Universe'', eds. Zimmermann,
H.U., Tr\"{u}mper, J., and Yorke H., 1996, MPE Report 263

\reference {} Cohen, J.G., Hogg, D.W., Pahre, M.A., Blandford, R., 1996,
ApJ, 462, L9

\reference {} Collins, C.A, Guzzo, G., Nichol, R.C., \& Lumsden, S.L.,
1995, MNRAS, 274, 1071

\reference {} Connolly, A.J., Csabai,I., Szalay,A.S., Koo,D.C.,
Kron,R.C. \& Munn,J.A., 1995, AJ, 110, 2655

\reference {} Cowie, L.L., Songaila, A., Hu, E.M. \& Cohen, J.G., 1996, AJ,
in press

\reference {} Fasano, G. \& Franceschini, A., 1987, MNRAS, 225, 155

\reference {} Geller, M.J. \& Huchra, J.P., 1989, Science, 246, 897

\reference {} Giovanelli, R. \& Haynes, M.P., 1993, AJ, 105, 1271

\reference {} Glazebrook, K., Ellis, R.S., Colless, M., Broadhurst, T.,
Allington-Smith, J. \& Tanvir, N., 1995, MNRAS, 273, 157

\reference {} Gronwall, C. \& Koo, D.C., 1995, ApJ, 440, L1

\reference {} Hughes J.P., Birkinshaw, M., Huchra, J.P., 1995, ApJ, 448,
L93

\reference {} Koo, D.C., 1981, ApJ, 251, 75

\reference {} Koo, D.C., 1985, AJ, 90, 418

\reference {} Koo, D.C., 1986, AJ, 311, 651

\reference {} Kron, R.G., 1980, ApJS, 43, 305

\reference {} Landy, S.D., Shectman, S.A., Lin, H., Kirshner, R.P.,
Oemler, A.A. \& Tucker, D., 1996, ApJ, 456, L1
 
\reference {} de Lapparent, V., Geller, M.J. \& Huchra, J.P., 1991, ApJ, 369, 273

\reference {} Le Fevre, O., Crampton, D., Hammer, F., Lilly, S. J., \&
Tresse, L., 1994, ApJ, 423, L89

\reference {} Liddle, A.R., Lyth, D.H., Schaefer, R.K., Shafi, Q. \&
Viana, P.T.P., MNRAS, in press

\reference {} Lilly, S.J., Le Fevre, O., Crampton, D., Hammer, F. \&
Tresse, L., 1995, ApJ, 455, 50

\reference {} Munn, J. A., Koo, D. C., Kron, R. G., Majewski, S. R., Bershady, M. A., \& Smetanka, J. J., 1995, ApJS, submitted

\reference {} Neff, S.G., Roberts, L. \& Hutchings, J.B., 1995, ApJS, 99,
349


\reference  {} Nichol, R.C., Holden, B.P., Romer, A.K., Burke, D.J., Collins, 
C.A., 1996, submitted to \apj


\reference {} Peacock, J.A., 1983, MNRAS, 202, 615

\reference {} Ramella, M., Geller, M.J. \& Huchra, J.P., 1992, ApJ, 384, 396

\reference {} Struble, M.F. \& Peebles, P.J.E., 1985, AJ, 90, 592

\reference {} Szalay, A.S., Ellis, R.S., Koo, D.C. \& Broadhurst, T.J.,
1991, in Primordial Nucleosynthesis and Evolution of Early Universe,
ed. K. Sato, J. Audouze (Kluwer, Japan), 435

\reference {} Ulmer, M.P., McMillan, S.L.W. \& Kowalski, M.P., 1989, ApJ,
338, 711

\reference {} West, M.J., 1989, ApJ, 344, 535

\reference {} Willmer, C. N. A., Koo., D.C., Ellman, N., Kurtz, M.J. \&  Szalay, A.S., 1996, ApJS, in press

\reference {} Zabludoff, A.I., Geller, M.J., Huchra, J.P. \& Ramella, M.,
1993, AJ, 106, 1301

\end{references}
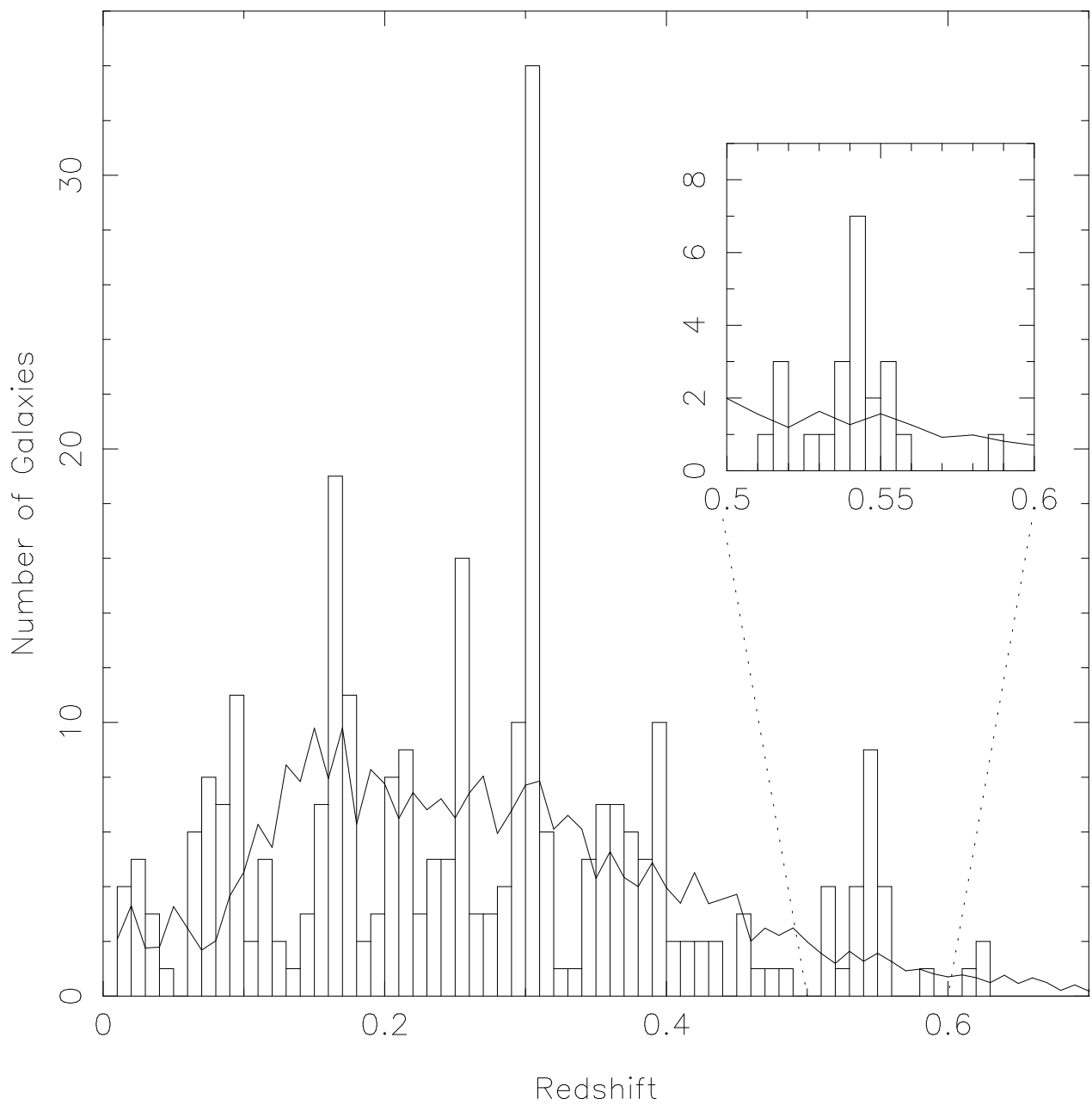
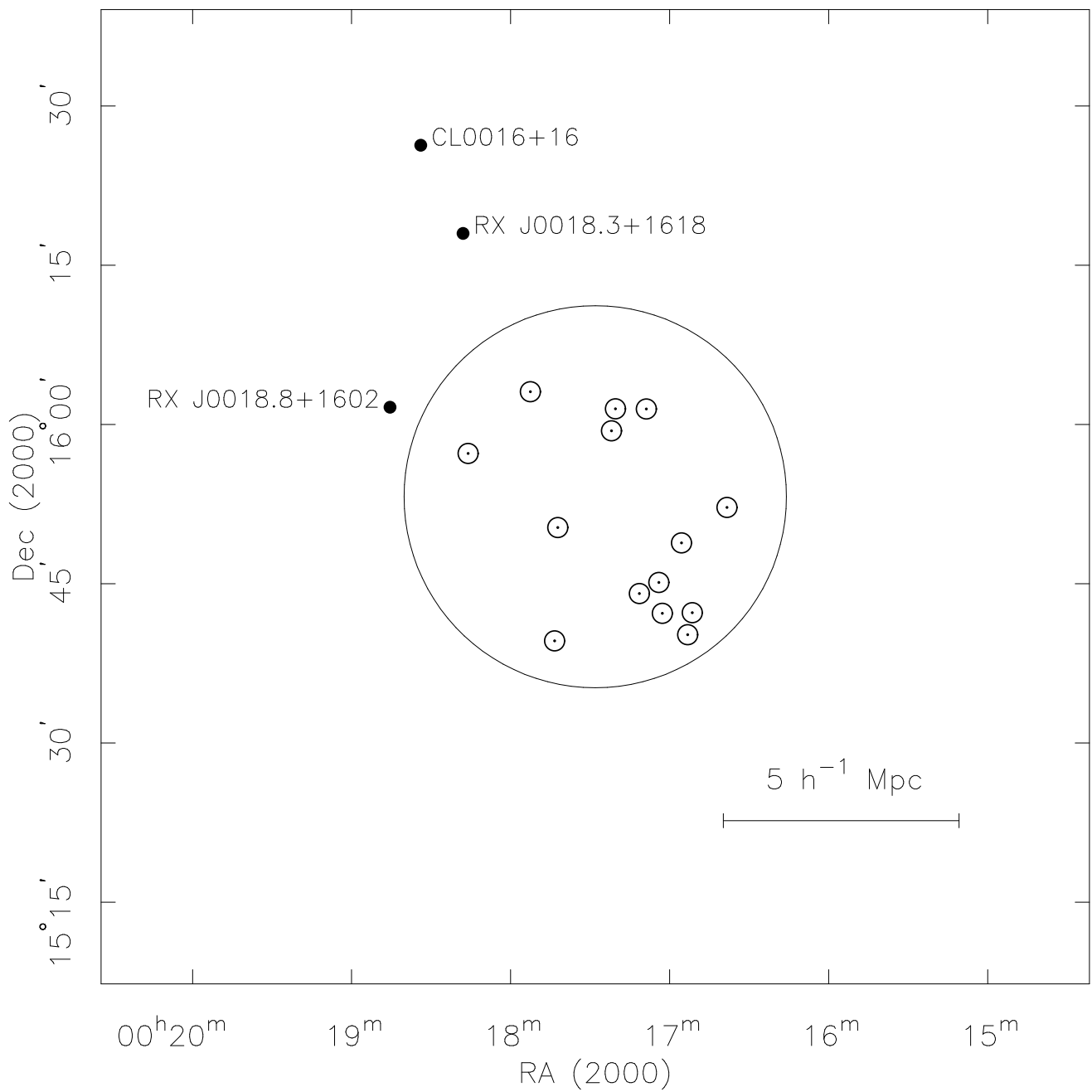
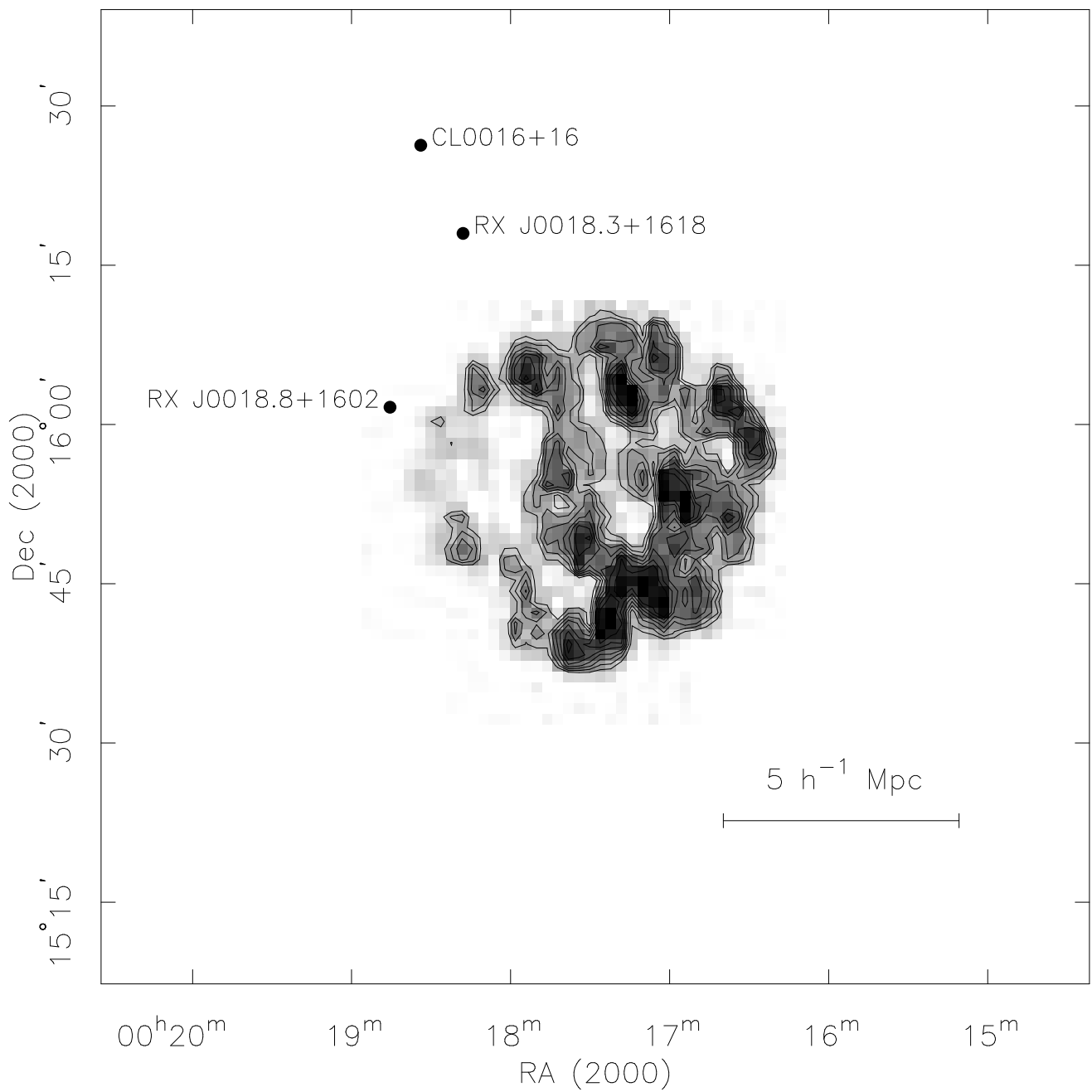
\end{document}